\documentclass[aps,12pt]{revtex4-1}

\usepackage{amssymb}
\usepackage[fleqn]{amsmath}
\usepackage{graphicx}
\usepackage{bm}

\newcommand{\dd}{\mathrm{d}}

\newcommand{\ke}[1]{\vert #1  \rangle}

\usepackage{xcolor}

\begin{document}

\title{X-Ray sum frequency generation; direct imaging of ultrafast electron dynamics} 

\author{J\'er\'emy R. Rouxel}
\email{jrouxel@uci.edu}
\author{Markus Kowalewski}
\author{Kochise Bennett}
\author{Shaul Mukamel}
\email{mukamel@uci.edu}
\affiliation{Department of Chemistry and Department of Physics and Astronomy, University of California,Irvine, CA 92697, USA}
\date{\today}

\begin{abstract}
X-ray diffraction from molecules in the ground state produces an image of their charge density, and time-resolved X-ray diffraction can thus monitor the motion
of the nuclei. However, the density change of excited valence electrons upon optical excitation can barely be
monitored with regular diffraction techniques due to the overwhelming background contribution of the core electrons. 
We present a nonlinear X-ray technique made possible by novel free electron laser sources,
which provides a spatial electron density
image of valence electron excitations. The technique,
sum frequency generation carried out with a visible pump
and a broadband X-ray diffraction pulse, yields snapshots of the transition charge densities, which represent the electron density variations upon optical excitation.
The technique is illustrated by ab initio simulations of transition charge density imaging for the optically induced electronic dynamics in a donor/acceptor substituted stilbene.
\end{abstract}

\maketitle

X-ray diffraction has been used for over a century to determine the structure of molecular crystals.
The experimental acquisition of time-dependent charge density movies by high resolution diffraction is now a reality thanks to recent advances in intense femtosecond X-ray free electron lasers 
(XFELs) \cite{emma2010first,Ding09prstab,natphotChini,miller2014femtosecond}
and tabletop ultrafast electron diffraction sources \cite{Miller14ar,Baum09cp}.
The static ground state charge density is commonly probed \cite{suryanarayana2013x}, with steady progress to tackle various difficulties, like the crystallization of large biomolecules and the phase recovery of the signal \cite{taylor2003phase}, yielding the molecular structure by revealing the location of the nuclei.
Time-resolved X-ray diffraction \cite{bennett2014time,schotte2003watching,glownia2016self} can provide stroboscopic snapshots of time-evolving excited state charge densities. 
The resulting real-time movies
 \cite{miller2010making,dixit2012imaging,vrakking2012x} monitor the optically triggered evolution of the molecular geometry.
Optical excitations typically involve few valence electrons (e.g., excitation from the highest occupied to the lowest unoccupied molecular orbital).
The ground and excited state charge densities are thus very similar and X-ray diffraction is dominated by the highly localized atomic core electrons. Despite this difficulty,  variations of few electrons over a strong background have been reported experimentally\cite{dixit2012imaging,vrakking2012x}.

In this paper, we propose a technique that directly images the \emph{change} in the charge density upon optical excitation and is thus particularly sensitive to the optically active electrons.
The technique can simultaneously monitor the rearrangement of the nuclei and the valence electrons in a photochemical reaction.
It offers the direct observation of transition charge densities (TCDs), which contribute to time-resolved diffraction when the molecule is initially prepared in a superposition of states. 
Our derivation for the diffraction image is based on the minimal coupling field-matter interaction Hamiltonian \cite{salammolQED}:
\begin{equation}
H_{\text{int}}(t) = - \int \dd\bold r \ \bold j(\bold r) \cdot \bold A(\bold r,t) + \frac{e}{2mc} \int \dd\bold r \ \sigma (\bold r) \bold A^2(\bold r,t)
\label{hint}
\end{equation}
where $\bold j(\bold r)$ and $\sigma(\bold r)$ are the current and charge density operators and $\bold A$ is the vector potential. The $\sigma (\bold r) \bold A^2(\bold r,t)$ term is
responsible for the off resonant diffraction as it is commonly used for X-Ray structure
determination.
The charge density is given by
\begin{align}
\sigma_{ij}(\mathbf{r_1}) = N \int d\mathbf{r}_2 \dots d\mathbf{r}_N \Psi_i(\mathbf{r}_1\dots\mathbf{r}_N) \Psi_j^*(\mathbf{r}_1\dots\mathbf{r}_N)
\end{align}
where $\Psi_{i/j}$ are the electronic eigenstates  and $\mathbf{r}_1..\mathbf{r}_N$ are the electronic coordinates.
The TCDs are the off diagonal elements, $\sigma_{ij} (i\neq j)$, which carry valuable chemical information about the molecular orbitals involved in the excitation (details can be found in supplementary materials).
When the ground and the excited states can each be described by a single Slater determinant, the TCD is given by the product of the two molecular orbitals differing between the two configurations.
The TCD thus provides a direct image of the electronic excitation and the location of the electron promotion.
Another way to view the TCD is as follows:
if we prepare a superposition of the excited state $e$ and ground state $g$, the expectation value of the charge density is given by a sum of $\sigma_{gg}$,
$\sigma_{ee}$ and the TCD $\sigma_{eg}$.
The latter thus represents an interference contribution to the charge density.
In the next section, we present the proposed technique in general terms. Then, we present short time electron dynamics simulations for a donor/acceptor molecule,
4-amino-4'nitrostilbene, which demonstrates how the TCD and time-evolving electron image of the valence electrons can be directly recovered. Finally, we discuss possible extensions in the conclusion.

\section*{Ultrafast sum-frequency X-ray diffraction}

\begin{figure}[h!]
\centering
\includegraphics[width=1.0\textwidth]{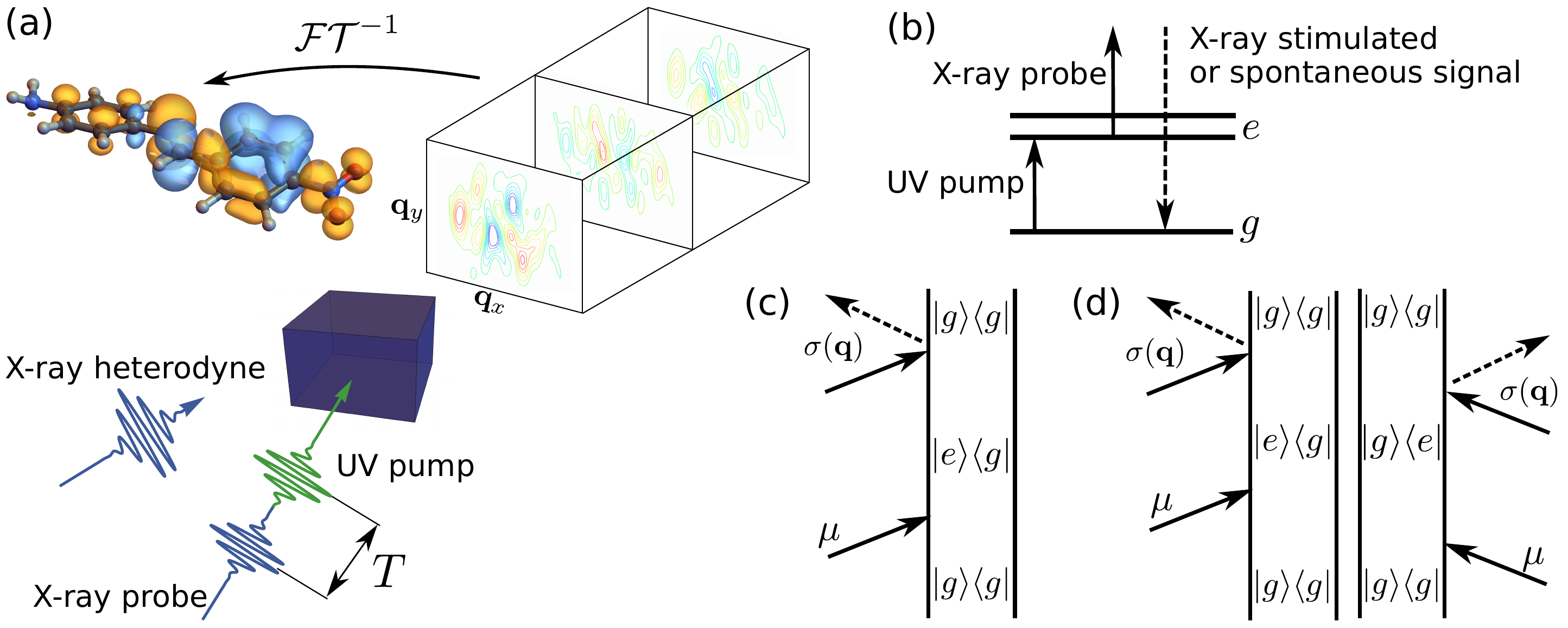}
\caption{Imaging $\sigma_{ge}(\bold q)$ by the SFG technique.
(a) Schematics of the TCD imaging process: A UV pump pulse creates a superposition
in the sample and an X-Ray probe pulse creates the diffraction pictures detected with a X-ray heterodyne pulse.
The TCD (upper left) can be reconstructed by an inverse Fourier transform.
Level scheme (b) for the SFG diffraction signals and corresponding double-sided diagrams for the heterodyne (c) and the homodyne (d) signals.
Note that the homodyne signal (d) stems from a two-molecule contribution which requires a long-range order in the sample.}
\label{fig:diagrams}
\end{figure}

The technique proposed here is a combined optical/X-ray nonlinear sum-frequency-generation (SFG) that provides images of electron dynamics through the TCDs  (see Fig.\ \ref{fig:diagrams}(a)).
This technique can also be seen as an anti-Stokes Raman scattering following a single interaction with an actinic pump. 
It is the lowest order nonlinear extension of time-resolved diffraction.

The proposed SFG technique, which records the $\sigma_{ge}(\bold q)$ image is laid out schematically in Figs.\ \ref{fig:diagrams}(a) and (b), combines a visible pump and an X-ray probe to study electronic coherences \cite{stiopkin2008heterodyne}. 
This is a direct analogue of the IR/visible setup commonly used to monitor vibrational coherences \cite{stiopkin2008heterodyne,shen1989surface}.
Time-domain SFG is routinely performed in the optical or infrared (IR) regime, most common in IR SFG from molecules on surfaces \cite{laaser2011time,Bethune:79}.
A visible/UV pulse first brings the molecule into a superposition of electronic ground state and excited states, which is then probed by a broadband hard X-ray pulse after a delay $T$.
Diagrams representing a time-dependent perturbation on the molecule density matrix \cite{mukamel1999principles} for the heterodyne and the homodyne detection schemes of SFG are sketched in Figs.\ \ref{fig:diagrams}(c) and (d).

\par
X-ray diffraction is commonly carried out in the spontaneous (homodyne) detection mode \cite{roslyak2010unified} where the signal is the diffraction image  \cite{yang2016diffractive}. 
The phase of the charge density in momentum $\bold q$-space is thus lost,
requiring to perform a phase retrieval algorithm \cite{taylor2003phase}.
Time-independent holographic stimulated (heterodyne) diffraction  \cite{Martin14nat,Tegze96nat} with a local reference oscillator can recover the phase of the scattered wave.
This common detection mode in the infrared and visible has been recently extended to the soft X-ray regime \cite{marchesini2008massively}. 
It requires an additional X-ray heterodyne pulse that interferes with the spontaneously emitted photons. 
This heterodyne pulse must be coincident with the X-ray probe pulse and relatively weak in order to be measured by an intensity detector.
Additionally, its phase must be controlled in order to recover the phase of the signal.
The scanning in momentum space can be done by rotating the sample and the heterodyne pulse (see supplementary materials) or by varying the spatial variation of the heterodyne pulse.

An interaction with a visible pulse first creates a valence electronic wavepacket.
If the pump selects a single electronic excited state, the signal provides a static image of a single TCD. 
However, when the pump creates a superposition of several  states,
the technique can provide images of the dynamics of electronic wavepackets. This adds valuable spectroscopic information to the structural information provided by ordinary diffraction.

Off-resonant diffraction processes are described by the $\sigma(\bold r) \bold A^2(\bold r)$ term in the minimal coupling field-matter interaction Hamiltonian, where $\sigma(\bold r)$ is the charge density operator and $\bold A(\bold r)$ is the vector potential of the radiation field.
In the heterodyne case, the emitted photon fields are superimposed with a classical field $\bm{\mathcal A_\text{het}}$, and the stimulated signal is defined as the field intensity in the $\bold k_\text{het}$ direction minus the intensity of the heterodyne pulse. Alternatively the experiment
can be done by imaging the coherent spontaneous emission and using a phase recovery.
The heterodyne detected diffraction image does not require
long range order in the sample, and can in principle be obtained with (oriented) molecules \cite{spence2004single,chatterley2017three}. 
The homodyne (spontaneous) detected diffraction image relies on intermolecular interference and requires a crystalline sample.
We focus on the heterodyne detected diffraction image
(see supplementary materials for derivation and discussion on the scaling of the signals) in the following.

The signal can be expressed in terms of a two-time correlation function of the charge density and dipole operator of the valence transition
\begin{equation}
S_{\text{het}}^{\text{SFG}}(\bold q,T)\propto\Im\int_{-\infty}^{+\infty} \dd t\int_{0}^{+\infty} \dd t_1
 \bold A_{\text{X}}(t) \cdot \bold A_{\text{het}}(t) \bold A_{\text{pump}}(t-t_1+T)
\langle\sigma(\bold q,t)\bm\mu^\dagger(t-t_1)\rangle
\label{stimulatedSFGfinal}
\end{equation}
where $\bold q =\bold k_\text{het}-\bold k_\text{x}$ is the momentum transfer,
$\bold A_{\text{X}}$ is the X-ray probe, $\bold A_{\text{het}}$ is the
heterodyne reference pulse, 
$\bold A_{\text{pump}}$ is the pump-pulse, and $\bm\mu = \int d\bold r \bold j(\bold r)$ is the dipole operator given by the integrated current density. Using integrated current densities is equivalent to invoking the electric dipole approximation for the pump interaction. 
Note that the heterodyne detected intensities are
interferences measured relative to the probe beam and thus can become negative, depending on the phase of $\sigma_{ge}(\mathbf{q})$.
Expanding Eq.\ \ref{stimulatedSFGfinal} in electronic eigenstates yields
\begin{equation}
S_{\text{het}}^{\text{SFG}}(\bold q,T)\propto\Im  \sum_e f_{eg}(T) \sigma_{ge}(\bold q) \bm \epsilon_{\text{pump}}\cdot\bm\mu_{eg}
\label{heterodyneSIG}
\end{equation}
where the lineshape function $f_{eg}(T)$ which contains the integrated pulse envelopes is given in supplementary materials. 
This function induces coherences between the ground state $g$ and excited states $e$ as permitted by pump the bandwidth.
The X-ray probe-pulse and the heterodyne reference pulse need to be shifted
in energy corresponding (see Fig.\ \ref{fig:diagrams}(b)) to the molecular
valence excitation and are required to be phase stable with respect to each other.

The spontaneous coherent signal can be written as a modulus square of an amplitude
(see supplementary materials, Eq. 34), making it not sensitive to the phase of the
X-ray probe-pulse.
The heterodyne detected image contains the same structural information as their spontaneous counterparts but resolve the phase problem since they give $\sigma_{ge}(\bold q)$ itself rather than its modulus square.
The stimulated signal intensities are stronger than their spontaneous counterparts.

\section*{Monitoring electronic dynamics via transition charge densities}

\begin{figure}[h!]
\centering
\includegraphics[width=0.8\textwidth]{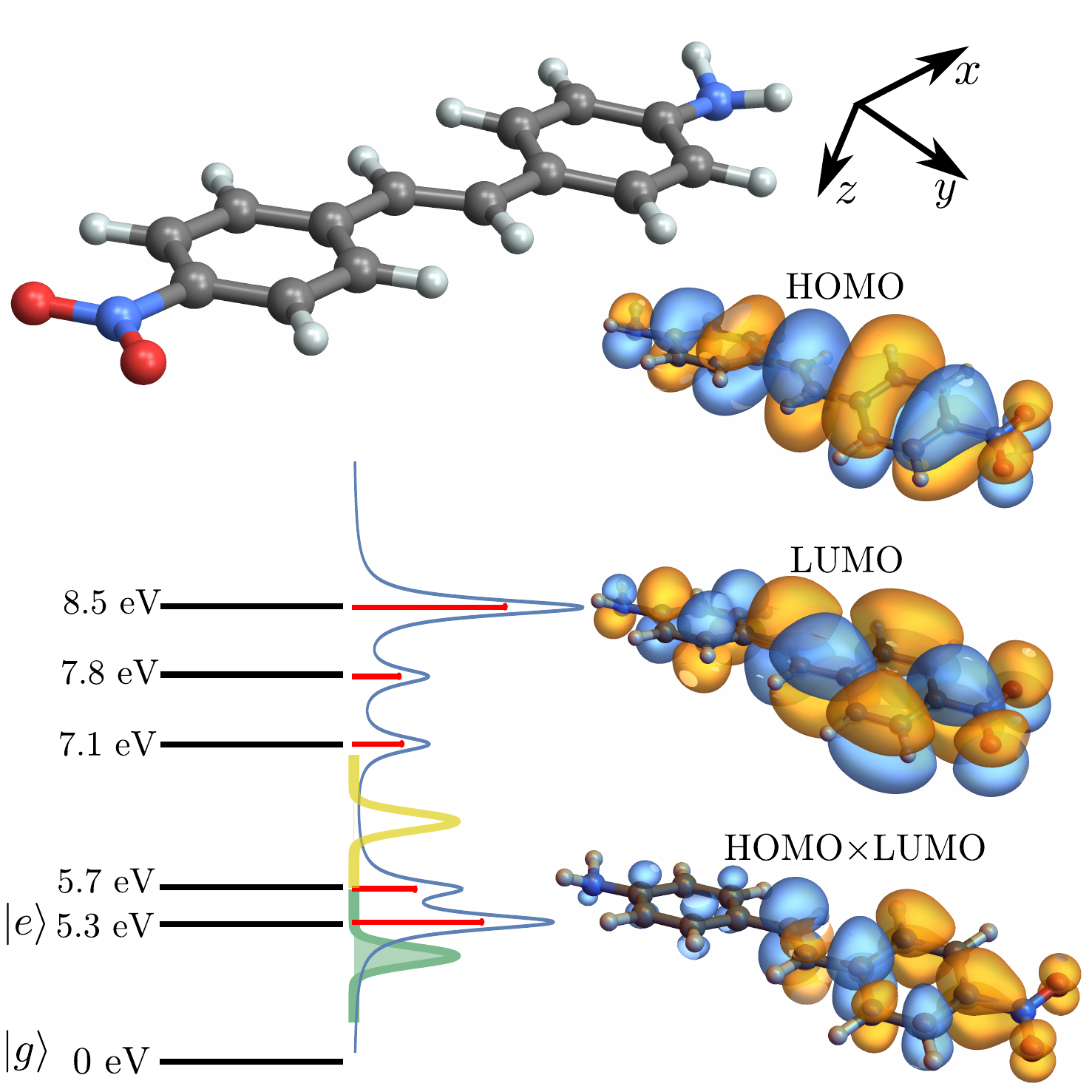}
\caption{Top : 4-amino-4'-nitrostilbene experiencing electronic dynamics probed by SFG time-resolved diffraction. The color scheme for the atoms is: carbon (gray), hydrogen (white), nitrogen (blue), oxygen (red). Bottom left:
electronic eigenstates including the electronic ground state $\ke{g}$, first excited state $\ke{e}$, and a set of multiple excited states. The linear absorption spectrum
is indicated by the blue curve and the red ticks. The pulses bandwidths are overlaid on the linear absorption spectrum and are centered at 3.68\,eV (green) and 6.47\,eV (yellow) for the single and multiple states preparations respectively. The highest occupied molecular orbital (HOMO) and and the lowest occupied molecular orbital (LUMO) are shown in the right panel (isovalue of 0.01). The product of HOMO and LUMO (lower right) approximately resembles the TCD and the corresponding SFG diffraction image (isovalue of $10^{-3}$).}
\label{fig:DAmol}
\end{figure}

We have calculated the stimulated SFG diffraction signals for 4-oriented amino-4'nitrostilbene
(Fig.\ \ref{fig:DAmol}). The TCDs and the integrated transition current densities in
Eq.\ \ref{heterodyneSIG} are obtained from ab initio calculation at the CASSCF(4/5)/6-31G* level of theory.
We work in the short time (few femtosecond) regime where we can neglect nuclear dynamics and 
radiation damage \cite{Amin17jpcb}.
The time evolution between the pump and the probe pulse is then determined solely by the electron dynamics.

\begin{figure}[h!]
\centering
\includegraphics[width=0.8\textwidth]{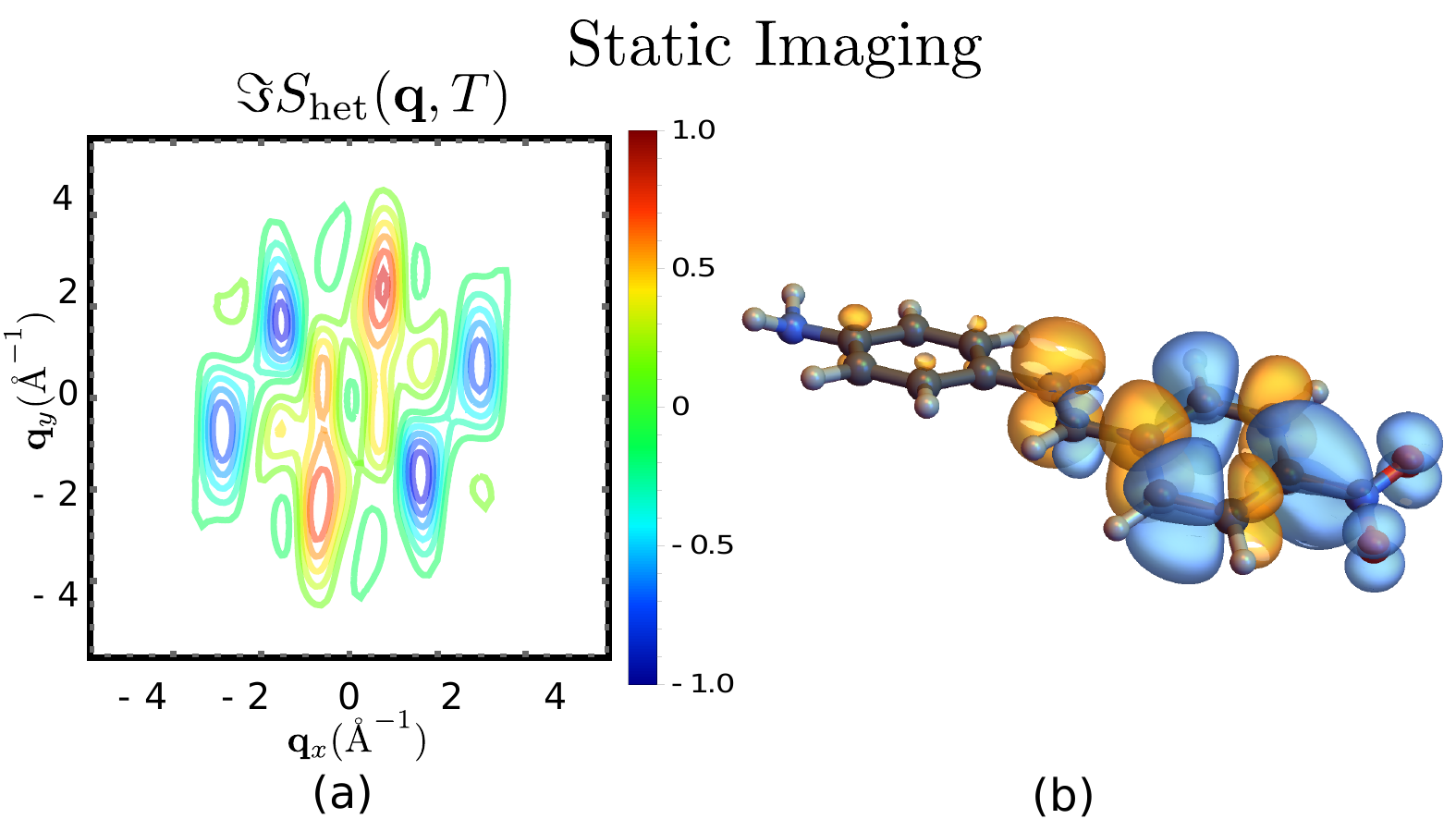}
\caption{Scattering of an X-ray pulse from oriented molecules pumped by a UV pulse that selects the first electronic excited state. (a) SFG diffraction pattern, calculated from Eq.\ \ref{heterodyneSIG}, in the $\bold q_z=0$ plane. (b) TCD image obtained by a Fourier transform to real space.
\label{fig3}}
\end{figure}

Fig.\ \ref{fig3} depicts the static imaging diffraction SFG patterns, Eq.\ref{heterodyneSIG}, in the $x-y$ plane (see Fig.\ \ref{fig:DAmol} for the scattering geometry).
The $y$-polarized pump-pulse (Fourier-transform limited with temporal spread (width) $\sigma = 5$\,fs and frequency $\omega=3.68$\,eV) creates a superposition between $g$ and a single excited state $e$. 
A 2\,fs off-resonant X-ray probe pulse is used to interrogate the superposition.
The heterodyne diffraction pattern in Fig.\ \ref{fig3}(a) carries the phase information of the TCD. 
The 3D SFG diffraction pattern can be Fourier transformed back into its real-space representation shown in Fig.\ \ref{fig3}(b), which corresponds to the TCD. 
The signal oscillation period corresponds to the difference between the pump and the matter transition frequencies. 
The shape of the signal, i.e. $\sigma_{ge}(\mathbf{r})$, closely resembles a product of the HOMO and LUMO as can be seen from comparison with Fig.\ \ref{fig:DAmol}.

\begin{figure}[h!]
\centering
\includegraphics[width=0.75\textwidth]{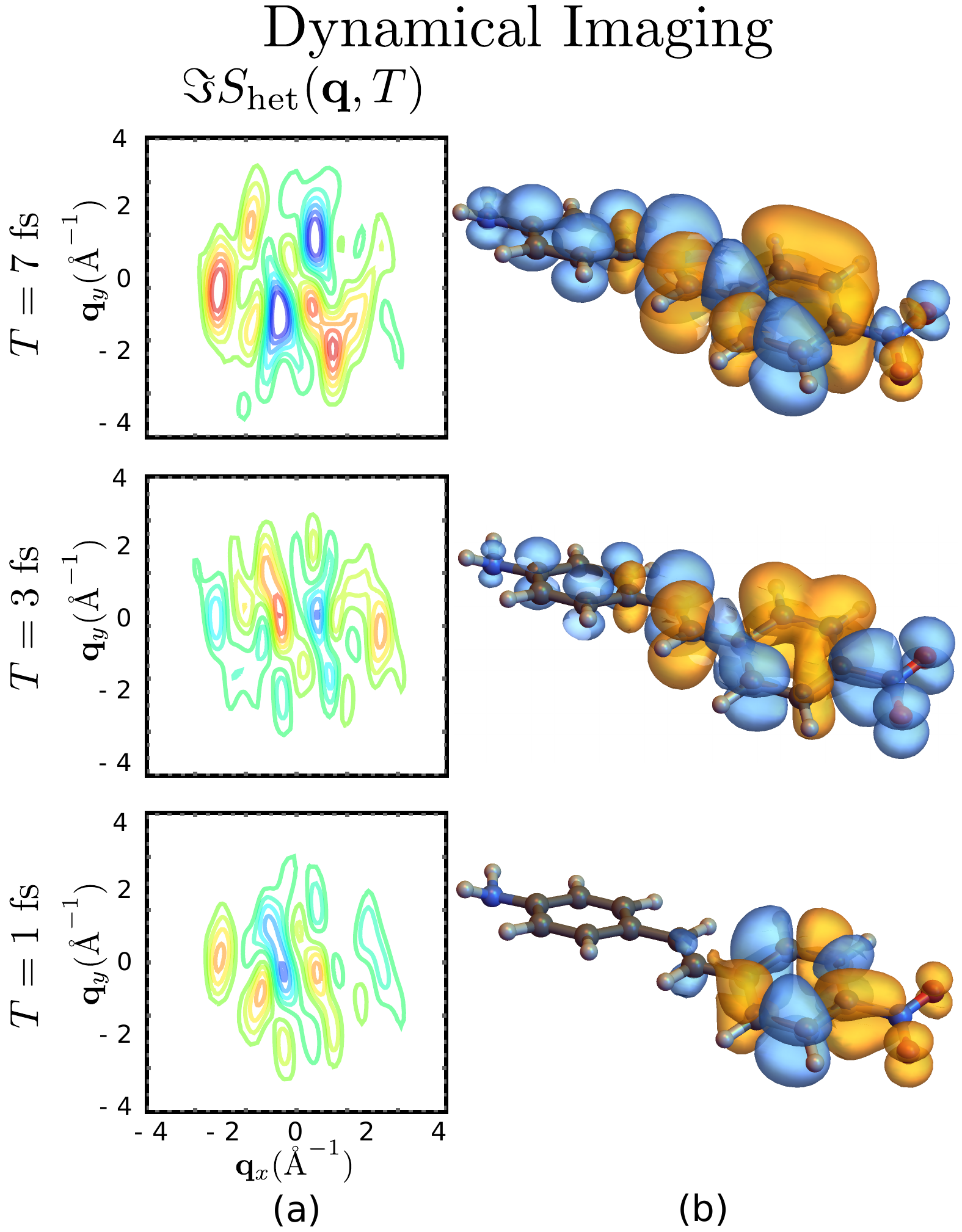}
\caption{Same as Fig.\ \ref{fig3} but the pump pulse a superposition of electronic excited states. (a) SFG diffraction patterns, calculated from Eq.\ \ref{heterodyneSIG}, in the $\bold q_z=0$ plane at various delays. (b) TCD images obtained by a Fourier transform to real space.
\label{fig4}}
\end{figure}

Dynamical imaging follows the evolution of an electronic wavepacket (superposition of excited state). A movie of this evolution is available in supplementary materials.
Few snapshots of this movie are displayed in Fig.\ \ref{fig4}.
Here, a UV pump-pulse (temporal spread (width) $\sigma = 5$\,fs and frequency $\omega=6.47$\,eV)
creates an electron wavepacket  composed of three excited states, as displayed in Fig.\ \ref{fig:DAmol}.
The signal is dominated by three transition matrix elements of the charge density operator : $\sigma_{1g}$ has only a minor contribution to the motion, $\sigma_{2g}$ which is mainly localized near the NO$_2$ (acceptor) group and $\sigma_{3g}$ also contains contribution in the NH$_2$ (donor) group.
Initially, we observe a linear superposition of comparable weight between $\sigma_{2g}$ and $\sigma_{3g}$ but as the time evolves, we see a beating between the two. For example, the signal at $T=$5.5 fs is almost uniquely a contribution for $\sigma_{3g}$ while at $T=$7.5 fs, $\sigma_{2g}$ strongly dominates the signal.
The time-resolved diffraction pattern then carries dynamical information through the TCDs $\sigma_{ge}$ between $g$ and the set of excited states $e$.\\

\section*{Conclusions}

We have demonstrated the capacity of time-resolved SFG diffraction to image the promotion of an electron into an excited state orbit. The signal provides access to an important matter quantity, the TCDs, and its heterodyne detected version can circumvent the phase problem.
The SFG diffraction patterns directly reveal the TCD, which can be interpreted as the quantum interference term between the ground and excited state charge densities.
When the pump selectively excites a single state, the diffraction pattern directly reveals a single TCD matrix element $\sigma_{ge}(\bold q)$, which carries information on the excited state orbitals in the corresponding valence excited state.
When an electronic wavepacket is prepared by a superposition of several states, the TCD carries both dynamical and structural information.
Ground state diffraction, in contrast, only monitors diagonal charge density matrix element $\sigma_{gg}$. Indirect reconstruction of orbital shapes
from high harmonic spectroscopy \cite{worner2010following}, photoelectrons
\cite{Villeneuve17sci} and time-resolved scanning tunneling microscopy \cite{Cocker16nat} has been reported. SFG diffraction in contrast gives direct access to the electron density.
Such experiments should be feasible in a near future \cite{sun2018current}: XFEL pulses possess the brilliance and the time duration necessary to detect nonlinear processes and electronic wavepacket evolution. The necessary phase stability may be achieved by
seeded FELs \cite{bencivenga2016four}.

The proposed technique can be extended in various ways.
First, scanning the pump/probe delay $T$ can be used to monitor molecular dynamical processes. 
The pump can launch nuclear dynamics and the time-resolved X-ray signal then reveals how the valence excited-to-ground TCD evolves in time, provided the pulse is short enough. The necessary X-ray pulses can be generated by existing XFEL sources \cite{Huang14prac}. The proposed technique require stable pulses at the edge of experimental capabilities
\cite{cao2016noncollinear,glover2012x,
bencivenga2016four,bencivenga2015four}, providing a path for X-ray sources improvement.

Second, the visible pump used in this work can be replaced by an X-ray pulse that creates valence electronic coherences through a stimulated Raman process \cite{mukamelSXRS1}. 
This should offer a much broader excitation bandwidth and higher time resolution.
Diffraction experiments are usually performed on ordered samples (crystals, oriented molecules) \cite{kupper2014x,chapman2011femtosecond,zhou2016x}. 
Using heterodyne detection of the diffraction image puts more constraints
on the X-Ray probe pulses but in return delivers the phase of the charge density.
Alternative to physically scanning the reference beam, spatial control
of the pulse phase could be explored in the future \cite{gu2004spatial,zhu2014arbitrary} combinded with phase cycling methods as they
are used in non-linear spectroscopy with \cite{Yan09cp} optical pulses.

Third, in a liquid or gas phase sample, some structural information is lost upon rotational averaging \cite{worner2010following,Villeneuve17sci}.
However, the present diffraction scheme could still yield valuable information. The visible and the Raman excitations are of different order in the exciting fields. In an isotropic sample, the Raman (odd-order $\chi^{(3)}$) signal vanishes while the (even-order $\chi^{(2)}$) SFG one does not, making the latter a new probe for time-resolved chirality \cite{1508.02890}. This will require extending the present work to include orientational averaging.
Finally, by frequency dispersing the probe and repeating the acquisition for multiple delays $T$, one can record a 3D $\bold q$-$\omega$-$T$ signal, revealing state selective spatial information when the system undergoes a complex electron and nuclear dynamics.

\section*{Acknowledgments}
The support of the Chemical Sciences, Geosciences, and Biosciences division, Office of Basic Energy Sciences, Office of Science, U.S. Deparment of Energy through Award No. DE-FG02-04ER15571 and of the National Science Foundation (Grant No CHE-1663822) is gratefully acknowledged. K.B. was supported by DOE.

\bibliographystyle{Science}
\bibliography{draft2}

\end{document}